\documentclass[12pt]{iopart}

\usepackage{amssymb}
\usepackage{graphics,graphicx}% Include figure files
\newcommand{\beq}{\begin{equation}}
\newcommand{\eeq}{\end{equation}}

\newcommand{\bea}{\begin{eqnarray}}
\newcommand{\eea}{\end{eqnarray}}

\newcommand{\ba}[1]{\begin{array}{#1}}
\newcommand{\ea}{\end{array}}
\newcommand{\ben}{\begin{enumerate}}
\newcommand{\een}{\end{enumerate}}
\newcommand{\bit}{\begin{itemize}}
\newcommand{\eit}{\end{itemize}}
\newcommand{\bde}{\begin{description}}
\newcommand{\ede}{\end{description}}

\newcommand{\kin}[1]{$k_{\mbox{\scriptsize in}}=#1$}

\newcommand{\kout}[1]{$k_{\mbox{\scriptsize out}}=#1$}
\newcommand{\koutt}{k_{\mbox{\scriptsize out}}}

\begin{document}

\title{Information transfer in community structured multiplex networks}

\author{Albert Sol\'e-Ribalta, Clara Granell, Sergio G\'omez and Alex Arenas}

\address{Departament d'Enginyeria Inform\`atica i Matem\`atiques, Universitat Rovira i Virgili, 43007 Tarragona (Spain)}

\begin{abstract}
The study of complex networks that account for different types of interactions has become a subject of interest in the last few years, specially because its representational power in the description of users interactions in diverse online social platforms (Facebook, Twitter, Instagram, etc.). The mathematical description of these interacting networks has been coined under the name of multilayer networks, where each layer accounts for a type of interaction. It has been shown that diffusive processes on top of these networks present a phenomenology that cannot be explained by the naive superposition of single layer diffusive phenomena but require the whole structure of interconnected layers. Nevertheless, the description of diffusive phenomena on multilayer networks has obviated the fact that social networks have strong mesoscopic structure represented by different communities of individuals driven by common interests, or any other social aspect. In this work, we study the transfer of information in multilayer networks with community structure. The final goal is to understand and quantify, if the existence of well-defined community structure at the level of individual layers, together with the multilayer structure of the whole network, enhances or deteriorates the diffusion of packets of information.
\end{abstract}

\maketitle

\section{Introduction}

% For Original Research Articles, Clinical Trial Articles, and Technology Reports the introduction should be succinct, with no subheadings.
%
% For Clinical Case Studies the Introduction should include symptoms at presentation, physical exams and lab results.
%
The study of transport properties of networks is becoming increasingly important due to the constantly growing amount of information and commodities being transferred through them. A particular focus of these studies is how to make the capacity of the diffusion of information in the network maximal while minimizing the delivery time. In the basic approach information is formed by units, the ``packets'', and the handling of information for processing and distribution takes finite time. Both network packet routing strategies and network topology play an essential role in networks' traffic flow. In realistic settings, like online social networks, the knowledge that any one has about the topology of the network is limited to its local area of influence. Consequently, much of the focus in recent studies has been on ``searchability'', the process of sending information to a target when the trajectory to reach the target is unknown. Moreover, given the limited capability of nodes to handle information packets and redistribute them, the problem of congestion arises \cite{guimera2001communication,arenas2004local,guimera2002optimal}. It has been observed, both in real world networks and in model communication networks, that the network flow collapses when the load (number of packets to be processed) is above a certain threshold \cite{guimera2002optimal}.

In general, most real and engineered systems include multiple subsystems and layers of connectivity, and it is important to take such features into account when trying to obtain a complete understanding of them. It is thus necessary to generalize the ``traditional'' network theory to multilayer systems in a comprehensive fashion \cite{kivela2014multilayer,Boccaletti20141}. Generally speaking, up to now, the description of networks so far has been developed using a single and combined snapshot of the connectivity, which is a reflection of instantaneous interactions or accumulated interactions in a certain time window. This description is limited when trying to understand the intricate variability of real complex systems, which contain many different time scales \cite{gauvin2014detecting} and coexisting structural patterns forming the real network of interactions \cite{mucha2010community}. This is the case of e-social networks that are constantly changing \cite{godoy2015long}, having some connections with very short lifetime and others that are persistent. Interest groups \cite{wellman1988social} are constantly being developed and growing, and individual nodes participate through different interests at the same time. An accurate description of such complexity should take into account these differences of interactions and their evolution through time. In the last couple of years, the scientific community on networks theory has focused on this issue and proposed a solution that has been commonly referred to as the multiplex network structure \cite{kivela2014multilayer}.

General flows on multiplex networks have also been in the focus of network scientists \cite{brummitt2012multiplexity, yaugan2012analysis, lambiotte2014random, lambiotte2015effect, bagnoli2015risk, zhao2014multiple, gomez2013diffusion, DeDomenico2014navigability}, and the consequences of having such topologies have been shown to be far from trivial. For example, in \cite{sole2013spectral} the authors found that a general diffusive process on top of the multiplex structure is able to speed up the less diffusive of the layers. It could also give rise to a super-diffusion process thus enhancing the diffusion of both layers. This striking result appears when the diffusion between the layers of the multiplex is faster than that occurring within each of the layers. These consequences are also observed in the discrete representation of diffusion by random walkers \cite{DeDomenico2014navigability}, and have explicit consequences on the navigability of the multiplex structure.

Here we fix our attention in the process of information transfer on top of multiplex networks. Specifically, we aim at determining the structural effects of a multiplex network endowing community structure, i.e.\ modular at each layer, on the dynamics of information transfer. To this end we have investigated a particular set up in which multiplex networks are built connecting different modular networks, and determining analytically the onset of congestion in the information flow. Our results reveal that when the community structure of the different layers is equivalent and the communities overlap, the multiplex offers higher resilience to congestion and consequently the system may improve information transfer compared to the individual layers. On the other side when the community structure is considerably different and communities still overlap the multiplex structure offers a balancing environment where the efficiency of the system is averaged. On the intermediate situation, that is community structure is similar in both layers and communities overlap, the effect of the multiplex structure is devastating and hinders information transfer by reducing the onset of congestion in the system.

%\begin{methods}
\section{Material \& Methods}

The proposed dynamical model considers that information flows through networks in atomic and discrete packets that are sent from an origin node to a destination node. Each node is an independent agent that can store as many packets as necessary. However, to have a realistic picture of communication we must assume that the nodes have a finite capacity to process and deliver packets. That is, a node will take longer to deliver two packets than just one. This physical constraint of the agents on delivering information can derive in network congestion. When the amount of information a particular agent receives is too large, it is not able to handle all the packets and some of them remain undelivered for extremely long periods of time. In this study, the interest is focused on when congestion occurs depending on the topology of the multiplex network, in particular, in relation to its community structure.

\subsection{Dynamics of information transfer}

The dynamics of the model is as follows. At each time step $t$, information packets are created at every node with rate $\rho$ (injection rate). Therefore, $\rho$ is the control parameter: small values of $\rho$ correspond to low density of packets and high values of $\rho$ correspond to the generation of a large amount of packets. When a new packet is created, a destination node, different from the origin, is chosen (uniformly) at randomly in the network. Thus, during the following time steps $t+1, t+2, \ldots, t+T$, the packet travels toward its destination. Once the packet reaches the destination node, it is delivered and disappears from the network.

The time that a packet remains in the network is related not only to the length of the path between the source and the target nodes, but also to the volume of packets that share its path. Nodes with high loads, i.e.\ high volume of accumulated packets, will take longer to deliver packets or, in other words, it will take more time steps for packets to cross regions of the network that are highly congested. We assume, without loss of generality, that nodes can  handle only one packet per time step (i.e.\, the delivery rate is $\tau = 1$), and undelivered packets are stored in a first-in-first-out queue attached to each node. The paths followed by packets between source and destination nodes are decided using a routing strategy, being shortest paths and random walks the most prominent strategies. It is important to note, however, that the model is not deterministic. For example, there may be several shortest paths between two nodes, one of them chosen randomly in the delivery of the corresponding packet. Moreover, the order in which packets are stored in the queues when several of them arrive in the same time step is undefined.
%Note that the existence of a multiplex structure allow us to define, in a somehow arbitrary way, how to choose origin and destinations. We decided to fix origin and destinations, as in the simple case of a one-layer network, i.e. nodes origin and destinations are chosen (uniformly) randomly regardless of the layer.

Previous work on single layer networks \cite{guimera2002optimal} shows that for low values of the injection rate of packets $\rho$ there is no accumulation of packets at any node in the network. Moreover, it is stated that the number of packets that arrive at node~$i$ is, on average, $\rho B_i / (N-1)$, where $B_i$ is the effective betweenness of node $i$ and $N$ the number of nodes in the network.
%It has been shown in previous work \cite{guimera2002optimal} concerning single layer networks that, for low values of the injection rate $\rho$, there is no accumulation of packets at any node in the network and the number of packets that arrive at node~$i$ is, on average, $\rho B_i / (N-1)$, where $B_i$ is the effective betweenness of node $i$ and $N$ the number of nodes in the network.
The effective betweenness is defined as the ratio between the number of paths that pass through node $i$, and the total number of paths traversing the network between any pair of nodes \cite{Newman2010Book}.

The onset of congestion is reached when a node receives more packets than it can deliver per time step, i.e.\ $\rho B_i / (N-1) > 1$. Therefore, the first node that collapses ($i^{\ast}$) is the one with largest effective betweenness ($B_{i^{\ast}}=\max_i(B_i)$), and the maximum injection rate for which the network is congestion free, the critical injection rate $\rho_c$, is given by
\begin{equation}
  \rho_c = \frac{N-1}{B_{i^{\ast}}}\,.
  \label{eq:rho_c_monoplex}
\end{equation}
The rest of the nodes will collapse with larger injection rates. However, up to now, it is not known how to analytically compute their critical injection rates since they not only depend on the topological betweenness but also on the overall network congestion.

In the generalization of the routing dynamics exposed to multiplex networks, the average number of packets arriving to node~$i$ in layer~$\alpha$ is $\rho L B_{i\alpha} / (N-1)$, where $L$ is the number of layers of the multiplex network and $B_{i\alpha}$ is the effective betweenness of node~$i$ in layer~$\alpha$. Thus, the critical injection rate also depends on the effective betweenness, which encapsulates the routing strategy and the topology of the network:
\begin{equation}
  \rho_c = \frac{(N-1)/L}{\displaystyle \max_{i,\alpha}(B_{i\alpha})}\,.
  \label{eq:rho_c_multiplex}
\end{equation}
Next, we extend the concept of betweenness to multiplex networks allowing the computation of the onset of congestion.

\subsection{Computation of betweenness in the multiplex}

The extension of any centrality measures to multiplex networks is not straightforward and requires special care. In many situations several extensions are possible and the choice of it may depend on the problem at hand. Many attempts have been done to extend single layer centrality measures to the multiplex framework \cite{Halu13MultiplexPageRank,battiston2014structuralMeasures,Sola2013Centrality}. Here, we follow the line described in \cite{deDomenico2015ranking}, which is mathematically grounded on the tensorial formalism for multilayer networks \cite{DeDomenicoPRX2013}.

\begin{figure}[!t]
	\centering
	  \includegraphics[width=8cm]{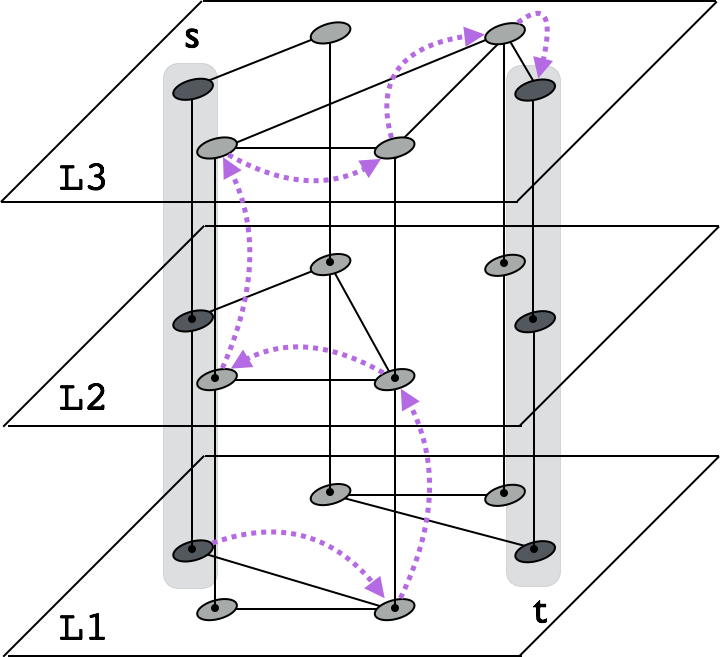}
	\caption{Example of a walk (dotted trajectory) between two individuals $s$ and $t$ using a multilayer network. The walker can jump between nodes within the same layer, or it might switch to another layer. This illustration evidences how multilayer structure allows a walker to move between nodes that belong to different (disconnected) components on a given layer (L1).}
        \label{fig:mplex}
\end{figure}

We start by defining a walk between two individuals $s$ and $t$ on a multiplex network as a sequence of nodes, following intralayer and/or interlayer edges, which starts at node $s$ ``in any layer'' and finishes in node $t$ ``in any layer''. Note that in this definition we do not care about the layer, just the node. The reasoning behind this lack of discrimination is that, in the multiplex structure, the different node replicas in the different layers correspond to the same individual (social networks) or location (transportation networks), thus it is only important to know if the packet has arrived, but not in which layer. %We only require that packets travel from the starting to the target individual or location.
Fig.~\ref{fig:mplex} shows an example of a walk between two nodes in a multilayer network where non-trivial effects can be observed because of the presence of interlayer connections that affect the navigation through the networked system \cite{DeDomenico2014navigability}.

Given the definition of a walk in the multiplex topology, the effective betweenness of a node~$i$ in layer $\alpha$, $B_{i\alpha}$, can be directly obtained as the fraction of walks that pass through node~$i$ in layer~$\alpha$ for every possible origin-destination pair $(s,t)$. In some cases it might be convenient to obtain the betweenness of node~$i$ irrespective of the layer. In this case, the betweenness can be obtained just by accumulation of the individual contributions of each layer where $i$ is represented, $B_i = \sum_{\alpha} B_{i\alpha}$.

For the specific case of the \emph{shortest path betweenness}, every walk is restricted to be the path with minimum distance that starts from the source node $s$ in any layer, and reaches the destination node $t$ in any layer. The distance function may take into account the weights of the edges the path traverses. In this work, without loss of generality, we assume the edges' weights are unitary and define the distance as the number of traversed edges in the path. The shortest path between two locations may be degenerated and consequently the set of shortest paths may contain paths using a single layer (classical shortest paths) and paths which change layer (pure multiplex paths). For an accurate computation of the shortest path betweenness special care must be taken with the path degeneration. A good and efficient algorithm can be found in \cite{Sole2014CentralityRankings}.

Equivalently to the shortest path betweenness, the \emph{random walk betweenness} depends on the particular definition of the network traversal procedure. In this case, a random walk is defined as a walk in which, at each time step, the next visited node is chosen with uniform random probability among the neighbors of the last visited node. The random walk betweenness is usually computed considering a transition matrix obtained from the adjacency matrix of the network. For a detailed description of random walks in a multiplex network, see \cite{DeDomenico2014navigability}. In this document we will use the classic random walk definition. For the random walk betweenness the walk degeneracy is enormous and consequently is impractical to compute the betweenness accounting for all the possible individual random walks. Fortunately, the random walk betweenness can be efficiently computed using matrix inversion and absorbing random walks \cite{Sole2015RandomWalkCentrality}.

\subsection{Community structure in multiplex}

Networks representative of complex systems are characterized by having community structure, meaning the presence of dense groups of nodes with sparse connections between them \cite{Fortunato201075}. It is known that dynamical processes running on top of such networks have a big dependency on community structure, which affects the process either by fostering or hampering it \cite{Liu05,Arenas06Syncro,Arenas200627}. As evidenced in several works \cite{bennett2015detection, mucha2010community}, when the different layers of a multiplex network exhibit community structure, the influence on the overall system is not trivial to determine.

Here, to uncover the basic effects of communities in information flow process, we propose a simple setting with imposed community structure where communities fully overlap between layers and the degree of each node of the network is kept constant. Each multiplex network consists of two layers, and each layer has 256 nodes distributed in four communities (64 nodes per community) \cite{newmanprebench}. The links are generated in such a way that the density of links inside the communities is always higher than the density between them. The networks are generated independently for each layer, resulting in a two-layer multiplex network with different community structure in each layer.

\begin{figure*}[!tb]
  \begin{center}
	 \begin{tabular}{p{0.3\columnwidth} p{0.3\columnwidth} p{0.3\columnwidth} c}

		\centering{$\mu = 0.03$} & \centering{$\mu = 0.23$} & \centering{$\mu = 0.37$} & \\
%%		\centering{$\mu_{\mbox{\superpetit{L2}}} = 0.03$} & \centering{$\mu_{\mbox{\superpetit{L2}}} = 0.23$} & \centering{$\mu_{\mbox{\superpetit{L2}}} = 0.37$} & \\
		\includegraphics[width=0.33\columnwidth]{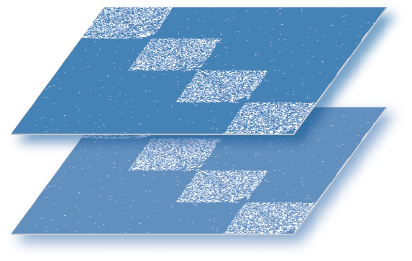}  &
		\includegraphics[width=0.33\columnwidth]{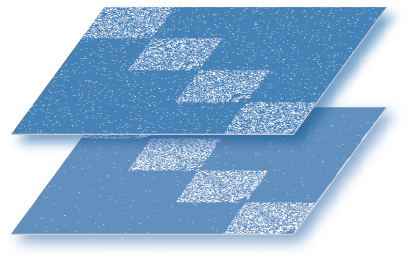}  &
		\includegraphics[width=0.3555\columnwidth]{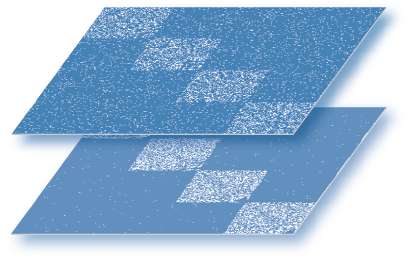} \\
		%\centering{$\mu_{\mbox{\superpetit{L2}}} = 0.03$} & \centering{$\mu_{\mbox{\superpetit{L2}}} = 0.23$} & \centering{$\mu_{\mbox{\superpetit{L2}}} = 0.37$} \\
		%\centering{$\mu_{\mbox{ L2}} = 0.03$} & \centering{$\mu = 0.23$} & \centering{$\mu = 0.37$} \\
		%\pbox{4cm}{first line \\ second line} & Second & Third \\
 	 \end{tabular}
	\end{center}
	\caption{Samples of the multiplex networks generated, represented by means of their superposed adjacency matrices. From left to right, the top layer diffuses its community structure (increasing mixing parameter), while the bottom layer remains fixed.}
	\label{fig:sketch}
\end{figure*}

For the experiments, we consider 12 different multiplex community structures and 300 different realizations for each. For all of them, we fix the bottom layer (L1) to \kin{31} and \kout{1} (i.e., 31~edges inside the community and 1~link outside, per node), which displays strong and clear communities, and we vary the community structure of the top layer (L2), which ranges from the previous strong block structure to a more diluted one (\kin{20} and \kout{12}) where the communities are almost imperceptible. We quantify the strength of the community structure of the L2 layer using a mixing parameter defined as $\mu = \koutt / \langle k \rangle$. Figure~\ref{fig:sketch} depicts three examples of such generated networks.

\section{Results}

To evaluate the influence of the multiplex networks with community structure in information transfer we assess several aspects of the information transfer dynamics, namely the shortest paths distribution, the packets ingoing rate of each node and the critical injection rate of the network.

\begin{figure}[!tb]
  \centering
    \includegraphics[width=\textwidth]{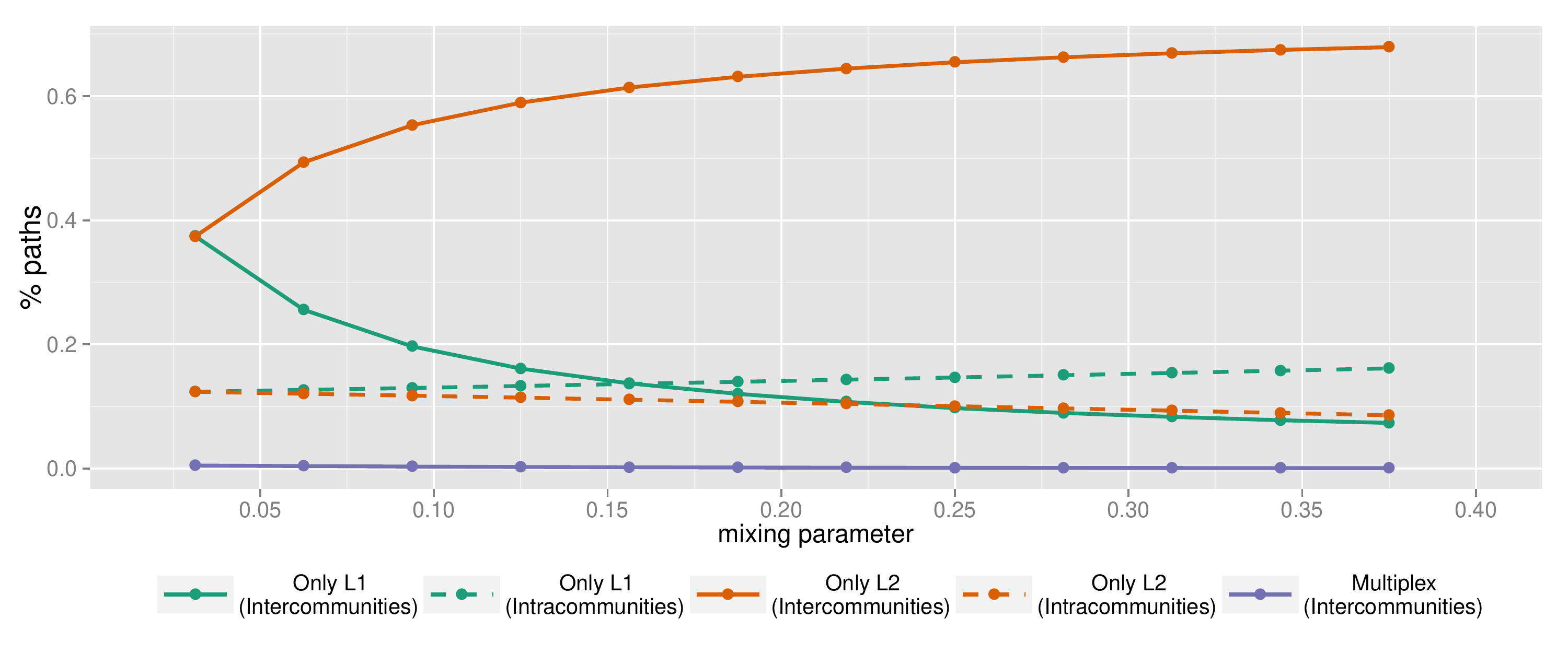}
  \caption{Shortest paths distribution in a multiplex network with community structure as a function of the mixing parameter. The plot shows the fraction of shortest paths that traverse the network using only layer~1 (with fixed topology), using only layer~2 (with varying topology) and using the full multiplex structure. The horizontal axis corresponds to the mixing parameter. For the paths that only use a single layer, we divide the contribution between paths where the source and target nodes are within the same community and in different communities. There are no intracommunity paths that use both layers.}
  \label{fig:path_distribution}
\end{figure}

Figure~\ref{fig:path_distribution} shows the obtained distribution of shortest paths in the different layers of the multiplex. In the case of having equivalent community structure in both layers (leftmost points in the plot), the multiplex structure provides a very good load balance where the same fraction of paths traverse using layer~1 and~2. In general, we can conclude that the effect of the multiplex is negligible for the overall system behavior since only a very small fraction of paths (0.5\%) makes use of the full multiplex structure. In fact, paths using both layers of the multiplex are only used in the case where the origin and destination are in different communities. As we increase the mixing parameter of the second layer, its community structure dilutes, enhancing the communication between communities but slightly hindering the transfer of information internally. This effect is evident in Fig.~\ref{fig:path_distribution}, which shows a large increase of intercommunity trips in the second layer and a small increase of intracommunity paths in the first layer. At the same time, the improvement of intercommunity paths in the more diffuse layer yields a disappearance of the (already small number of) shortest paths using both layers.

To assess the microscopic behavior of the system we show how the ingoing rate of packets to each node varies with respect to the mixing parameter. We compute the ingoing rate of each node of the multiplex structure as
\begin{equation}
	\hat{\sigma}_{i\alpha} = \frac{L~B_{i\alpha}}{N-1}.
\end{equation}

\begin{figure*}[!tb]
  \begin{center}
	 \begin{tabular}{ll}
	 	{\bf A} & {\bf B} \\
		\includegraphics[width=0.45\columnwidth]{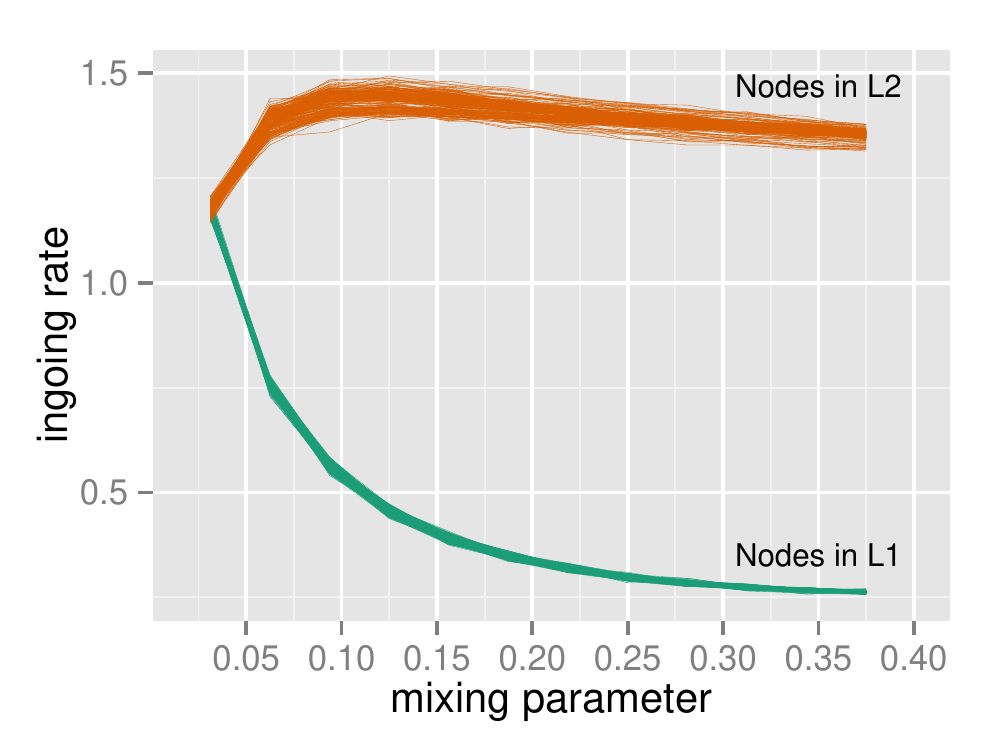}  &
		\includegraphics[width=0.45\columnwidth]{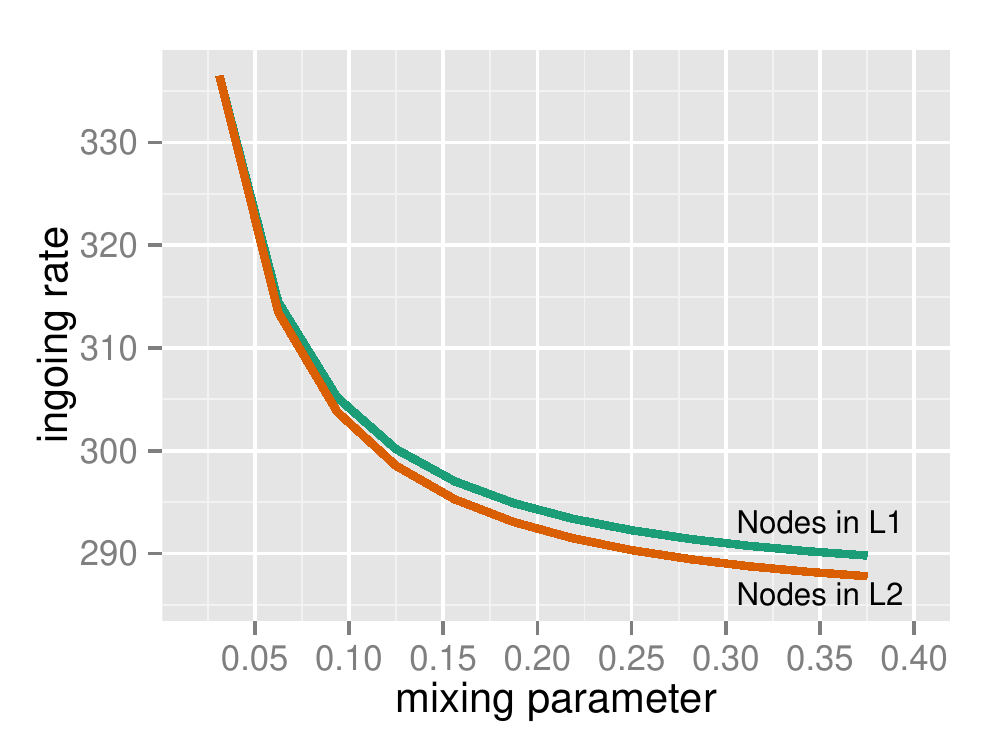} \\
 	 \end{tabular}
	\end{center}
	\caption{Ingoing rate of each node in the network for different mixing parameter values, and two routing strategies: shortest paths ({\bf A}) and random walks ({\bf B}). The different colours indicate the layer to which the node belongs to.}
	\label{fig:ingoing_rates}
\end{figure*}

Results are shown in Fig.~\ref{fig:ingoing_rates}. For the shortest path routing strategy (subplot A) we observe a clear distinction between the behavior of nodes in layers~1 and~2. As can be seen in Fig.~\ref{fig:path_distribution}, the main effect on the increasing of the mixing parameter is clearly a migration of the shortest paths from layer~1 to layer~2, i.e. paths that traversed layer~1 now find a more efficient route through layer 2, which has a more diluted community structure. This migration of paths should increase the ingoing rate of nodes in layer~2 similarly to the observed decrease of ingoing packets of layer~1. This is the situation for small mixing parameters, but increasing the mixing parameter means also an increase in the efficiency of layer~2 routing packets between nodes in different communities, which in turn substantially reduces the overall node betweenness provoking an interesting tradeoff that will prescribe the final efficiency of the full system. These two opposed effects (migration of shortest paths and reduction of node betweenness) have a huge impact on the ingoing rate of nodes in layer~2, which experience a constant decrease after the maximum ingoing rate is reached. For the random routing strategy the scenario is completely different. The increase of the mixing parameter has an equivalent impact in both layers, which experience an important decrease of the ingoing rate.

%but as we keep increasing the mixing parameter we also increase the efficiency of layer~2 routing packets between intercommunity nodes, which has a tremendous effect in the overall reduction node betweenness. These two opposed effects (migration of shortest paths and reduction of node betweenness) have a huge impact on the ingoing rate of nodes in layer~2, which experience a constant decrease after the maximum ingoing rate is reached. For the random routing strategy the scenario is completely different. The increase of the mixing parameter has an equivalent impact in both layers, which experience an important decrease of the ingoing rate.

\begin{figure}[!tb]
  \centering
    \includegraphics[width=\textwidth]{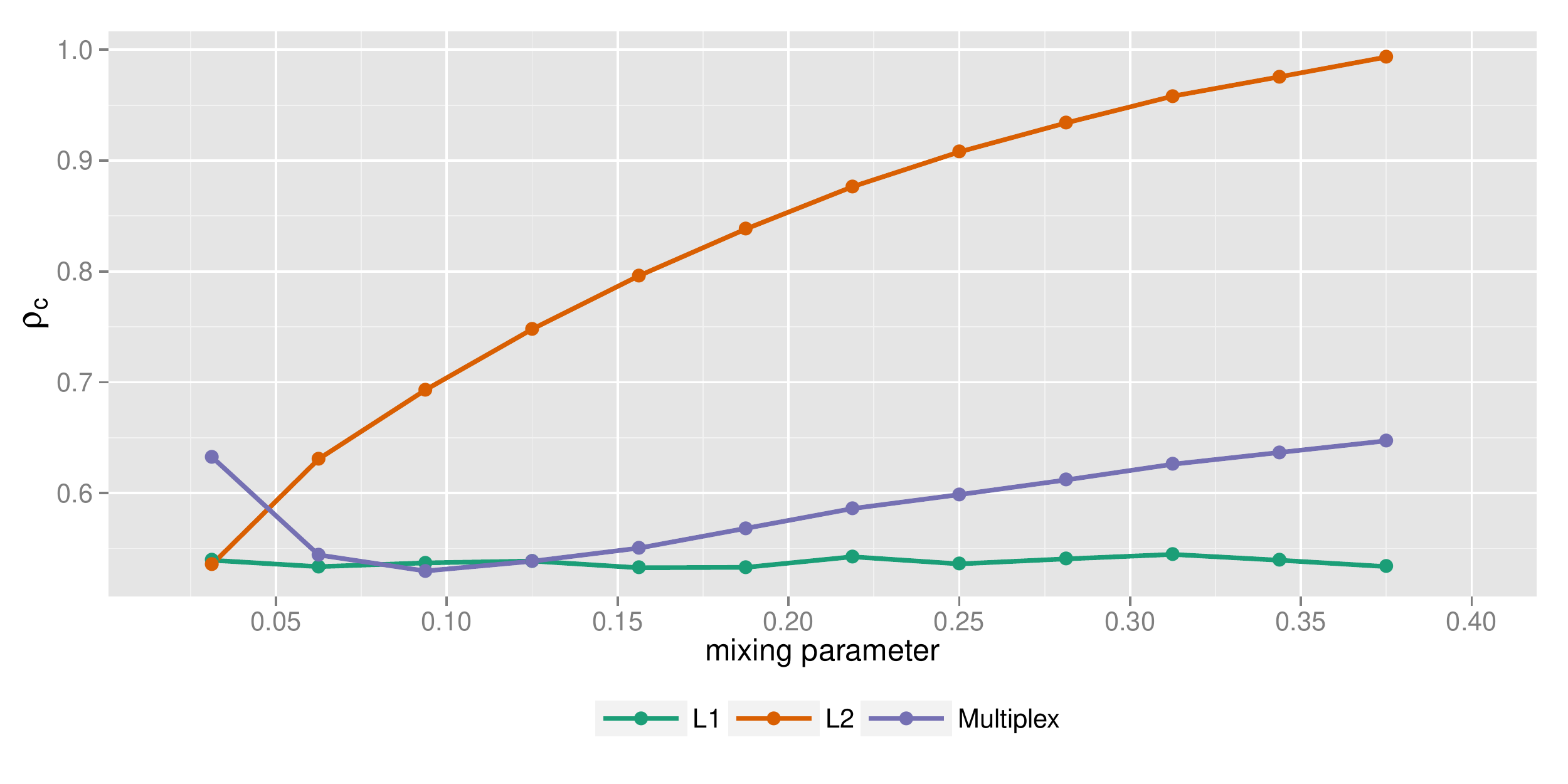}
  \caption{Comparison of the critical injection rate of layer~1, layer~2 and the multiplex for different mixing parameter values and shortest path routing strategy. The values of the critical injection rate for single layer network and multiplex networks are computed using Eqs.~(\ref{eq:rho_c_monoplex}) and~(\ref{eq:rho_c_multiplex}) respectively.}
  \label{fig:rho_c_SP}
\end{figure}

\begin{figure}[!tb]
  \centering
    \includegraphics[width=\textwidth]{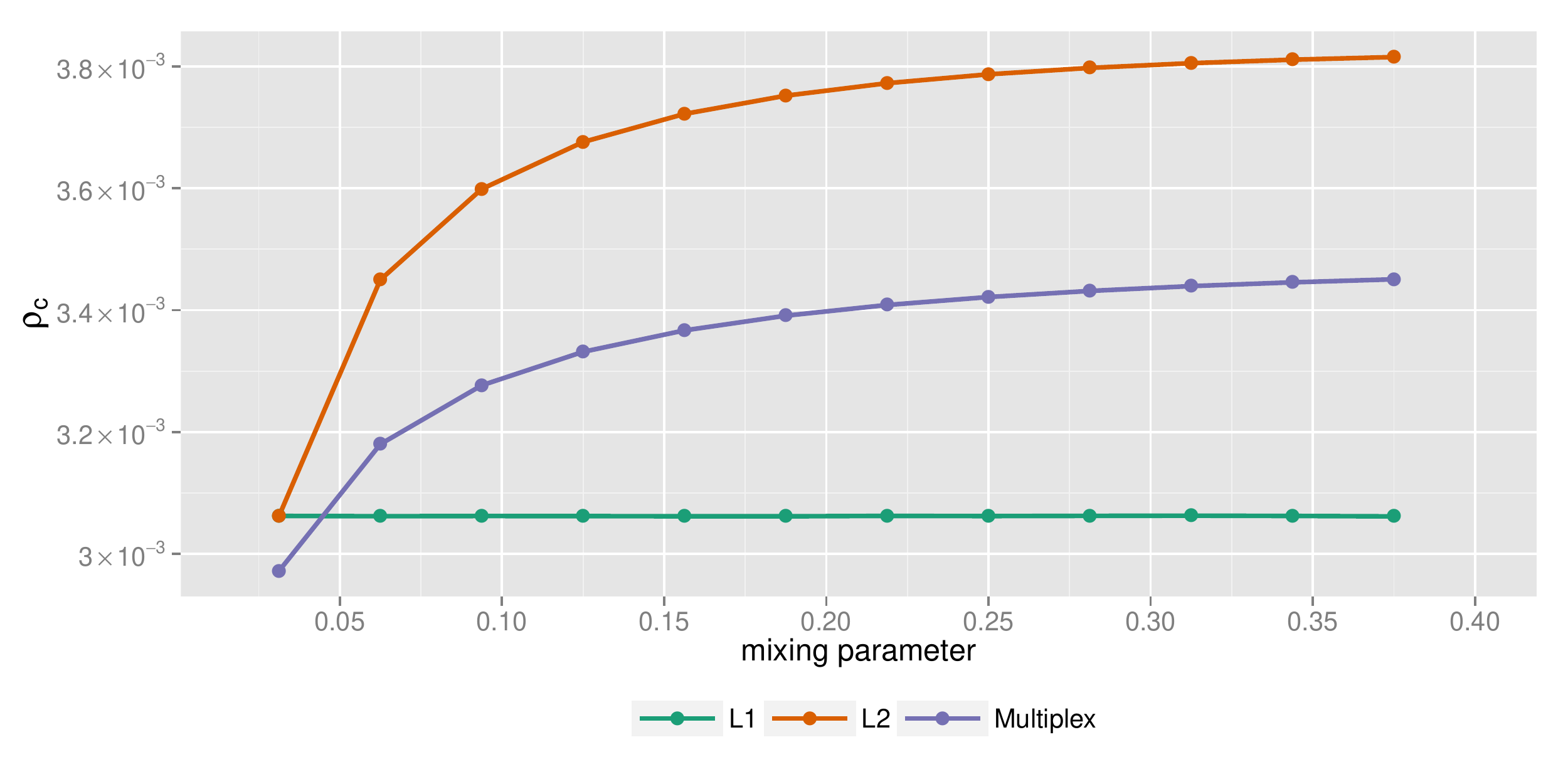}
  \caption{Comparison of the critical injection rate of layer~1, layer~2 and the multiplex for different mixing parameter values and random walk routing strategy. The values of the critical injection rate for single layer network and multiplex networks are computed using Eqs.~(\ref{eq:rho_c_monoplex}) and~(\ref{eq:rho_c_multiplex}) respectively.}
  \label{fig:rho_c_RW}
\end{figure}

Figures~\ref{fig:rho_c_SP} and~\ref{fig:rho_c_RW} show the effect of the community structure on the critical injection rate $\rho_c$. For the shortest path routing strategy (Fig~\ref{fig:rho_c_SP}) the critical injection rate of the multiplex network reaches its minimum value around~$\mu=0.1$. This minimum indicates that there is a worst-case scenario for which the multiplex topology is less efficient than the individual layers. On the other side, the behavior of the critical injection rate of layer~2 is monotonically increasing. This situation is expected since a less clear community structure leads to a reduction of the average shortest path, which in turn is positively correlated with a decrease of the node betweenness.

In general, if we compare the values of $\rho_c$ for the multiplex network and the separate layers L1 and L2, we clearly observe three possible situations: (i) the multiplex is more resilient to congestion (efficient) than the individual layers. This situation arises when both layers have a similar community structure. (ii) The multiplex is less efficient than any of the layers. This setup corresponds to the minimum resilience of the multiplex structure. And (iii), the multiplex efficiency achieves a medium value which is a trade-off between the resilience of both layers. In a real setup, this situation would mean that joining those two layers in a multiplex improves the resilience observed in one layer at the cost of deteriorating the resilience observed in the second layer.
%This situation, in a real application, would mean that the multiplex improves the resilience observed in one layer at the cost of deteriorating the resilience observed in the second layer.
However, as we can observe in the plot, the reduction of the efficiency is larger than the average of the efficiency of both layers and consequently the coupling of layers is inefficient.
%the cost of improving the resilience of the first layer considerable.

With respect to the random walk routing strategy (Fig~\ref{fig:rho_c_RW}), the situation is completely different. For similar community structures the multiplex worsens the efficiency, because the paths get trapped within the communities. For different community structures, in general, we obtain a efficiency that corresponds to the average efficiency of both layers.

\section{Discussion}

We have shown that packet information flow can be compromised when community structure is considered in some layer of the multiplex network structure.
Since community structure implies the presence of topological bottlenecks, the information flows migrate to those layers where these constraints are relaxed (diluted communities).
We have shown that community structures produce a non-trivial effect in the transfer of information and in the resilience to information flow congestion, that here defines the efficiency of the structure.
Essentially, the better defined the communities, the more affected the packet transportation. Information tries to avoid bottlenecks and packets migrate towards the layer where the community structure is diluted, because it is more efficient, but as a direct consequence of this migration the most efficient layer becomes overloaded. This nonlinear relation makes the problem of assessing the performance of the multiplex structure particularly challenging.

Using the analytical approach presented, we are able to determine for any multiplex topology what is the onset of congestion in the information flow and how it compares with the onset of the individual layers. We have also provided results, for very specific scenarios, of shortest path and random walk routing strategies respectively. The results show that the shortest path approach heavily depend on the particular sharpness of the prescribed communities. This work provides the starting point for the discrete flow analysis of more complicated scenarios of community structure in multiplex networks.

\section*{Acknowledgments}

This work has been supported by Ministerio de Econom\'{\i}a y Competitividad (Grant FIS2012-38266) and European Comission FET-Proactive Projects PLEXMATH (Grant 317614). A.A.~also acknowledges partial financial support from the ICREA Academia and the James S. McDonnell Foundation.

%\bibliographystyle{frontiersinSCNS} % for Science, Engineering and Humanities and Social Sciences articles, for Humanities and Social Sciences articles please include page numbers in the in-text citations
%\bibliographystyle{frontiersinHLTH&FPHY} % for Health and Physics articles

%%%%%%%%%% Insert bibliography here %%%%%%%%%%%%%%
\section*{References}

%\bibliographystyle{unsrt}
%\bibliography{bibliography}

\end{document}